\documentclass[prd,superscriptaddress,amsfonts,amssymb,amsmath,showpacs,twocolumn]{revtex4-2}

\usepackage{bm}
\usepackage{amsfonts}
\usepackage{latexsym}
\usepackage{graphicx}
\usepackage{amsmath}
\usepackage{palatino}
\usepackage{mathpazo}
\usepackage{textcomp}
\usepackage{float}
\usepackage{booktabs}
\usepackage{dcolumn}
\usepackage{booktabs}
\usepackage{multirow}
\usepackage{hyperref}
\usepackage{orcidlink}
\hypersetup{colorlinks,citecolor=blue}
\usepackage[caption=false]{subfig}
\usepackage{xcolor}

\begin{document}
	\color{black}       
\title{	A Nonsingular Logarithmic Bouncing Cosmology in $f(R,T)$ Gravity with Thermodynamic Viability}
	\author{Anjani\orcidlink{0009-0000-1108-7648}}
	\email{anjaliguleria002@gmail.com}
	\affiliation{Department of Mathematics, Maharaja Agrasen University, Kalujhanda, Himachal Pradesh-174103, India} 
	
	\author{Pankaj Kumar\orcidlink{0000-0002-7156-5637}}
	\email{pankaj.11dtu@gmail.com}
	\affiliation{Department of Mathematics, Maharaja Agrasen University, Kalujhanda, Himachal Pradesh-174103, India}
	
	\author{ S. H. Shekh\orcidlink{0000-0002-1932-8431}}
	\email{da_salim@rediffmail.com}
	\affiliation{Department of Mathematics, S.P.M. Science and Gilani Arts, Commerce College, Ghatanji, Yavatmal, \\Maharashtra-445301, India.}
	\affiliation{Pacif Institute of Cosmology and Selfology (PICS), Sagara, Sambalpur, Odisha 768224, India}
		\affiliation{L N Gumilyov Eurasian National University, Astana 010008, Kazakhstan.}	
	
	\author{Milan Srivastava\orcidlink{0000-0002-7156-5637}}
	\email{milandtu@gmail.com}
	\affiliation{Department of Mathematics and Data Science, Sharda University, Greater Noida, India.}

	\begin{abstract}
\textbf{Abstract:} We present a nonsingular bouncing cosmological model in the framework of modified $f(R,T)$ gravity within a spatially flat Friedmann--Robertson--Walker universe. A logarithmic time-dependent scale factor is assumed to realize a smooth transition from a contracting phase to an expanding phase without encountering an initial singularity. Based on this assumption, the dynamical evolution of the Hubble parameter, deceleration parameter, energy density, and pressure is obtained for various choices of the model and the matter--geometry coupling parameter to confirm the occurrence of a successful bounce. The effective equation of state parameter is examined to characterize the cosmic fluid during different evolutionary phases. The violation of energy conditions, necessary for the realization of the bouncing behavior, is also discussed. The stability of the model is investigated using the squared speed of sound and is found to remain positive within the allowed parameter space, indicating classical stability. Furthermore, constraints on the matter--geometry coupling parameter are obtained by demanding positive energy density, negative pressure, and a viable cosmological evolution. The obtained cosmological constraint on the coupling parameter is also shown to be compatible with the currently available compact-object constraints. The thermodynamic behavior of the model is examined by testing the generalized second law of thermodynamics. The total entropy production rate remains negative during the contracting phase and changes its sign to positive during the expanding phase. However, it becomes singular at the bouncing point, reflecting the breakdown of the standard thermodynamic description during the transition phase.
	\end{abstract}
	\maketitle
\section{Introduction}
General Relativity (GR) is widely regarded as the most successful theory of gravitation, providing a fundamental geometrical description of the gravitational phenomena in spacetime \cite{Einstein1916, Will2014}. However, observations at both galactic and cosmological scales reveal significant discrepancies with GR predictions. In particular, the flat rotation curves of galaxies, supported by gravitational lensing and galaxy cluster observations, cannot be explained only by visible matter. On cosmological scales, the late-time accelerated expansion of the universe has been confirmed by various probes such as Type la supernova \cite{1,Per1}, Baryon Acoustic Oscillations (BAO) \cite{Eienstein}, large-scale structures formation \cite{Tegmark1}, high precision measurements of the Cosmic Microwave Background (CMB) from Wilkinson Microwave Anisotropy Probe (WMAP) and Planck missions \cite{DES,Planck,Spergel1}. To address these observations within GR, additional dark components-cold dark matter(CDM) to explain galactic-scale phenomena and the cosmological constant($\Lambda$), commonly referred to as dark energy(DE), to account for cosmic acceleration- are introduced in GR, forming the standard $\Lambda-CDM$ model. Despite its remarkable observational success, the $\Lambda-CDM$ model faces conceptual challenges. These include, the unknown physical nature of dark components, the fine tuning and cosmic coincidence problems, possible indications of evolving DE with time, the observational tensions among datasets and the presence of initial singularity \cite{constp,evol DE,inf,Sahni}. Such unresolved issues suggest the need for more general gravitational action beyond are required to describe the universe on large scales. Motivated by this, extensive efforts have been devoted to explore modified theories of gravity as alternatives to GR. These theories typically arise from modifications of the Einstein-Hilbert(EH) action. One of the earliest and simplest extension is the $f(R)$ theory of gravity, originally proposed in \cite{Buchd}, where gravitational dynamics are described from extensions in standard Riemannian geometry in which EH-action is generalized to an arbitrary function $f(R)$ of the Ricci scalar $R$ \cite{Duru,Kle,Bamba2008}. Further generalizations include higher-order curvature corrections such as $f(R,G)$ gravity, where $G$ is Gauss Bonnet invariant  \cite{frg,fgr2}, as well as torsion (teleparallel gravity) and non-metricity-based formulations, including $f(T)$, $f(T,B)$, where $T$ is the torsion and $B$ is the boundary term and symmetric teleparallel $f(Q)$ gravity \cite{ft,ft2,ft3,q}. Another important class of modified theories that involve the interaction between matter and spacetime geometry which give rise to additional effective gravitational terms beyond GR and can naturally generate negative pressure, is seen in $f(R,T)$ gravity, $f(Q,T)$ gravity, where $T$ represents the trace of energy momentum tensor \cite{fqt,fqt3,SK}. Although modified gravity successfully derives the late time acceleration, the physical origin of the cosmic expansion remains a significant open problem for both astrophysicists and cosmologist.

The early rapid inflationary expansion, driven by inflaton field, which addresses the horizon, flatness, large scale structure formation and relic density problems, indicates that $\Lambda$ can account only for current cosmic acceleration \cite{inf}. However, inflation is formulated within the framework of GR and therefore inherits initial conditions of the so-called big-bang singularity, where classic explanation of spacetime breaks down. Even within the inflationary cosmology, the existence of an initial spacetime singularity appears unavoidable as demonstrated by Hawking et. al singularity theorems \cite{HawkingPenrose}. Moreover, the exponential expansion and attractor nature of inflation erase the information of pre-inflationary universe \cite{BVG,Liddle}. Consequently, standard cosmology remain incomplete due to its reliance on unexplained components and initial singular conditions. These fundamental limitations encouraged the development of non-singular models such as bouncing cosmologies and quantum gravity inspired models which aim to provide a self consistent description of universe's evolution. 
In such models, the universe does not originate from an initial singularity; instead, it contracts, reaches a minimum size (the bounce), and then expands, a process that may occur as part of a single transition or in cyclic sequences. Prominent examples are Ekpyrotic or cyclic models (involving brane collisions in higher dimensional string theory), matter bounce scenarios, quantum corrections induced bounces and Gauss-Bonnet bounces \cite{Lohakare2022,Steinhardt2002}.\\
In the context of GR, Friedmann equations combined with Null Energy Conditions (NEC) forbid the transition from contraction to expansion as $\dot{H}\leq0$ and thus prevent the realization of a non singular cosmological bounce \cite{EC,Haro2017,boun,Nov}. Furthermore, Raychaudhuri equation implies that for matter satisfying $(\rho+p)\geq0$ the geodesics in GR are focused which leads to the singularities \cite{Ray}. In contrast, modified gravity theories offer several ground breaking developments such as produces early and late time acceleration without ($\Lambda$), allowing natural violation of NEC and admit non-singular bouncing solutions and the resolution of classical spacetime singularity. The comprehensive reviews summarize these developments across modified gravity theories in \cite{allr,allr2}.

In this work, we investigate the bouncing cosmology within the framework of $f(R,T)$ theory of gravity which is proposed by Harko et al. \cite{Harko}, where $R$ stands for Ricci curvature scalar and T represents the trace of energy momentum tensor. A special feature of $f(R,T)$ gravity is the dependence on $ T $ which may arise from imperfect fluids or quantum anomalies that can drive cosmic acceleration without DE \cite{Alv,Singh}. Controlled violations of energy conditions further constrained the functional form $f(R,T)$ and provided a base for stability \cite{consfunc,Houndjo a,Houndjo b}. The theory has also been tested with the observational data sets in \cite{TT,TT1} with the study indicating improved  consistency compared to $\Lambda-CDM$ model \cite{Bouli}. Stable bouncing and phantom like cosmological solutions within $f(R,T)$ has been constructed by various authors. Shabani et al. initiated stable bounce constructions, while Bamba et al. \cite{BH2} has achieved phantom divide line i.e $w= -1$, with effective geometry effects. The studies by Sahoo et al. and Bhattacharjee et al. provided detailed stability analyses \cite{BH1,BH3,BH4}. Singh et al. \cite{BH5} developed phantom like stable cosmologies consistent with observations and Agarwal et al. \cite{Agrawal} proposed physically viable matter bounce scenario. Although, this theory offers a richer cosmological effects and has attracted attention in recent years, it also faces some challenges. The non-conservation of $T$ deviates the motion of test particle from geodesic, while the ambiguities in the matter Lagrangian $\mathcal{L}_{m}$ affect the field equations. In addition, many models suffer from instabilities and negative sound speeds. These issues have been interpreted in terms of particle production and non-equilibrium thermodynamics \cite{thermo}. Recent theoretical developments have focused on extending the formalism through Palatini approaches, anisotropic cosmological models and generalized theories like $f(R,\mathcal L_{m},T)$ aiming to address stability, consistency and matter geometry coupling in a unified framework \cite{HarkoLobo,Nagpal an}.

In the present work, we introduce a logarithmic form of the scale factor to observe a non-singular bouncing universe. To the best of our knowledge, this logarithmic form have not been explored earlier to discuss bouncing cosmology in any gravitational theory. Our analysis is focused on the dynamical evolution of the model during different cosmic phases, including the non-singular bouncing regime. For a clear physical interpretation, we consider the contracting phase for $t<0$, the bounce occurring at $t=0$, and the expanding phase for $t>0$. We examine the corresponding dynamic and thermodynamic behavior of presented non-singular bouncing model. We also discuss classical stability and energy conditions of the bouncing model. To further assess the viability of the proposed model, the cosmological constraint obtained on the matter--geometry coupling parameter is also compared with the currently available compact-object constraints.

The remainder of the paper is organized as follows. Section~2 presents the field equations of $f(R,T)$ gravity. A brief overview of bouncing cosmology is given in Section~3. In Section~4, we introduce the logarithmic scale factor and derive the corresponding Hubble and deceleration parameters. The dynamical behavior of the model, including the equation of state parameter, energy conditions, and stability analysis, is presented in Section~5. Section~6 is devoted to the thermodynamic analysis of the model. Section~7 examines the consistency of the cosmological constraint on the matter--geometry coupling parameter with the currently available compact-object constraints. Finally, the main results and concluding remarks are presented in Sections~8 and~9, respectively.\\

\section{$f(R,T)$ Theory and Its Field Equations }
 The $f(R,T)$ theory is a generalization of $f(R)$ theory, where Harko et al. \cite{Harko} considered that the gravitational part depends not only on Ricci curvature scalar $R$ but also on $T$, where $T$ expressed as the trace of the energy-momentum tensor of matter i.e., $T = T_{\mu\nu}g^{\mu\nu}$.  For the  $f(R,T)$ theory, the form of EH action in the unit (set $8 \pi G = c = 1$) is given by
\begin{equation}\label{1}
S=\frac{1}{2}\int[f(R,T)+2\mathcal{L}_{m}]\sqrt{-g} d^4x ,
\end{equation}
where $\mathcal{L}_{m}$  is the matter Lagrangian density , g is  the determinant of the metric tensor $g_{\mu\nu}$ . The energy- momentum tensor, $T_{\mu\nu}$  is defined as
\begin{equation}\label{2}
T_{\mu\nu}= - \frac{2\delta ( \sqrt{-g} \mathcal {L}_{m})}{\sqrt{-g}\;\delta g^{\mu\nu}}.
\end{equation}
In $f(R,T)$ gravity the Lagrangian density $\mathcal L_{m}$ of matter depends on $g_{\mu\nu}$ only and not on its derivatives. Therefore, the above eq. (\ref{2}) becomes
\begin{equation}\label{3}
T_{\mu\nu}= g_{\mu\nu}\mathcal{L}_{m} - 2\frac{\partial\mathcal L_{m}}{\partial  g^{\mu\nu}}.
\end{equation}
An interesting consequence of this theory is non-conservation of divergence of energy - momentum tensor, i.e, $ \nabla^{\mu} T_{\mu\nu} \neq 0 $  which governs the possible exchange of energy between matter and geometry. This non-conservation can be interpreted thermodynamically as a mechanism for particle production in an open system dynamics \cite{thermo}.
The variation in action of the gravitational field S in eq. (\ref{1}) with respect to the metric tensor $g^{\mu\nu}$ yields the modified field equation  in metric form as
\begin{widetext}
\begin{equation}\label{4}
f_{R}(R,T) R_{\mu\nu}  -  \frac{1}{2} f(R,T) g_{\mu\nu} +  ( g_{\mu\nu} \Box  -  \nabla_{\mu} \nabla_{\nu}) f_{R}(R,T)  = T_{\mu\nu} -  f_{T}(R,T) T_{\mu\nu}  -  f_{T}(R,T) \Theta_{\mu\nu}
\end{equation}
\end{widetext}
where $  f_{R}  (R, T ) =  \frac { \partial  f (R, T )} { \partial R}$ and $ f_{T} (R, T ) =  \frac{ \partial  f (R, T) }{\partial  T}$, $\nabla_{\mu}$ denotes the covariant derivative w.r.t $ g_{\mu\nu}$ and $\Box$ is the D'Alembert operator. The term $ \Theta_{\mu\nu} $ present in the field equation characterizes the dependence of matter on metric and  can be written as
\begin{equation}\label{5}
\Theta_{\mu\nu} = -2T_{\mu\nu} + g_{\mu\nu}\mathcal L_{m} - 2g^{\alpha\beta} \frac{\partial^{2}\mathcal L_{m}}{\partial g^{\mu\nu}\partial g^{\alpha\beta}}.
\end{equation}
Generally, the gravitational field  in $f(R,T)$ is influenced by the physical nature of the matter source through the tensor $ \Theta_{\mu\nu}$. The functional freedom in $f(R,T)$ gravity of function f enables one to construct feasible, adaptable and different matter models in theoretical framework. The renowned general forms include
 $f(R,T) = \{R +f(T) ,  f_{1}(R) +f_{2}(T) ,  f_{1}(R)+f_{2}(R)f_{3}(T) \}$
such that $ f_{n} ( R ) $ and  $f_{n} ( T ) $ are arbitrary functions of $R$ and $T$ respectively. Each choice of these functional forms give rise to distinct dynamical equations and leads to diverse cosmological interpretations. Over here we are working on notable form $f(R,T) = R + f(T)$ , where the term $R$ is usual Einstein-Hilbert action and the additional correction  $f(T)$ modifies general relativity. This modification introduces an extra force that can account for dark matter like effects. It also generates effective cosmological constant that evolves dynamically with matter density. Therefore the eq. (\ref{4}) for the form  $f(R,T) = R + f(T)$ becomes
\begin{equation}\label{6}
 R_{\mu\nu} - \frac{1}{2}Rg_{\mu\nu} = T_{\mu\nu} - ( T_{\mu\nu}+ \Theta_{\mu\nu})f '(T) + \frac{1}{2}f(T) g_{\mu\nu}.
\end{equation}
Here, prime denotes the derivative w.r.t argument. The matter lagrangian $\mathcal L_{m} $ can be chosen as $\mathcal  L_{m} = -p $, where p being the thermodynamic pressure of matter content of the universe. Now eq. (\ref{5}) gives $ \Theta_{\mu\nu} = -2 T_{\mu\nu} -p g_{\mu\nu}$. Using this result, Eq. (\ref{6}) reduces to
\begin{equation}\label{7}
 R_{\mu\nu} - \frac{1}{2}Rg_{\mu\nu} =  T_{\mu\nu} - ( T_{\mu\nu}+ pg_{\mu\nu})f '(T) + \frac{1}{2}f(T) g_{\mu\nu}
\end{equation}
In this work, we assume a spatially homogenous and isotropic flat Friedmann - Lemaitre - Robertson - Walker (FLRW) space-time which is expressed in co-moving co-ordinates by the line element
\begin{equation}\label{8}
ds^2 = dt^2 - a^2(t) (dx^2 + dy^2 + dz^2).
\end{equation}
For the line element , the eq. (\ref{7}) gives the following the field equations
\begin{equation}\label{9}
3H^2 = \rho + (\rho + p)f '(T) + \frac{1}{2}f(T),
\end{equation}
\begin{equation}\label{10}
2\dot{H} + 3H^2 = -p +  \frac{1}{2}f(T).
\end{equation}
Here, $\rho$ and  $p$ denote the total content of density and pressure of the universe, respectively. We consider  $f(R,T)=R+f(T)$  with $f(T) = \lambda T$, $\lambda$ is a coupling parameter and  function $f (T)$ is modifying GR by generating non-minimal interaction between matter and geometry. Therefore, the field equations (\ref{9}) and (\ref{10}) becomes
\begin{equation}\label{11}
3H^{2}=\rho+\lambda(\rho+ p) + \frac{1}{2}\lambda T
\end{equation}
\begin{equation}\label{12}
2\dot{H} + 3H^{2}= - p +  \frac{1}{2}\lambda T
\end{equation}
Here, the overhead dot on H represents the derivative w.r.t.  cosmic time $t$. Using above two eqs. we have obtained expression for energy density and pressure as,
\begin{equation}\label{13}
\rho = \frac{3(\lambda+1)H^{2}-\lambda\dot{H}}{(2\lambda+1)(\lambda+1)}, \hspace{0.3cm}
p =- \frac{(3\lambda+2)\dot{H}+3(\lambda+1)H^{2}}{(2\lambda+1)(\lambda+1)}
\end{equation}

\section{Bouncing Cosmology}
The fascinating questions, what existed before the big-bang? and what preceded time itself, if big-bang marks the beginning of time?, lie at the heart of modern cosmology. According to GR, the big-bang corresponds to the singular point where the scale factor vanishes, and physical quantities such as energy density, pressure and curvature diverge. This breakdown highlights the need for alternative frameworks such as bouncing cosmologies in which the universe has a meaningful pre big-bang history. One such approach is provided by bouncing cosmologies where the universe transitions smoothly from a contracting phase to a finite minimum size known as the bounce, before re-entering an expanding phase. In this framework, the big-bang is replaced by a transition point rather than a true beginning. For a successful bounce scenario the following conditions must be satisfied \cite{N}:
\begin{itemize}
\item The scale factor $a(t)$ in contracting phase decreases with ($\dot{a}(t)<0$), while it grows steadily with cosmic time in expanding phase($\dot{a}(t)>0$). When $a(t)$ attains a non-zero minimum value, the model is defined as a non-singular bouncing model and at this point, the bounce, $\dot{a}(t) = 0$ and the acceleration is positive i.e ($\ddot{a}(t)>0$) which ensures the transition.
\item At the bouncing point, the Hubble parameter also vanishes ($H=\frac{\dot{a}}{a}=0$). In an FRW universe, $\dot{H}=-4\pi G(\rho+p)$ , $\rho_{eff}+p_{eff}<0$ which implies $\dot{H}>0$ must be finite and positive, this necessitates a temporary violation of NEC near the bouncing epoch. This is followed by $H<0$ and $H>0$ for contracting and expanding phases respectively.
\item The equation of state parameter ($EoS$) must cross the phantom divide line $(w < -1)$ in the neighbourhood of the bouncing point which facilitate the violation of energy conditions required for bounce.
\end{itemize}
In the next section, we introduce a form of scale factor and  analyze the behaviour of important cosmological parameters, i.e, Hubble, Deceleration and equation of state parameters in order to discuss the evolution of bouncing universe.
\section{Parameterization of Scale Factor}
Introduced through Friedmann field equations in Einstein's GR equation, the scale factor $a(t)$ plays a pivotal role in distinguishing among various cosmological scenarios. Each of these is characterized by distinct physical and mathematical conditions. Its time-dependent nature encodes different cosmic phases of the universe, including the inflationary era $a(t)\sim e^{Ht}$, the radiation dominated era $a(t)\sim t^{\frac{1}{2}}$ , which are strongly supported by observational evidence of CMB radiation and Type Ia supernovae data.  Several  authors have explored bouncing solutions by proposing the various forms of Hubble parameter or scale factor such as linear, anstaz, power law etc \cite{a,a1,a2}. Following this approach, we introduce a logarithmic form of scale factor as,
\begin{equation}\label{14}
 a(t) = m \log (\alpha + \beta t^{2})
\end{equation}
where $ m, \alpha $ and $ \beta $ are constants, $m,\beta>0$ and $\alpha>1$. For $t<0$, $\dot{a}(t)=\frac{2m\beta t}{\alpha+\beta t^2}<0 $, with corresponding $\ddot{a}(t)=-\frac{4m\beta^2 t^2}{(\alpha+\beta t^2)^2}+\frac{2m\beta}{\alpha+\beta t^2} <0 $ and in the vicinity of bouncing point $\ddot{a}(t)=-\frac{4m\beta^2 t^2}{(\alpha+\beta t^2)^2}+\frac{2m\beta}{\alpha+\beta t^2}>0 $. At $t=0$, the bounce point, the contraction slows down as its rate change vanishes $\dot{a}(t)=0$, with positive acceleration $\ddot{a}(t)=\frac{2m\beta}{\alpha}>0$. At this instant, $a(t)$ acquires finite positive value without reaching zero, i.e, $a(t)=m\log\alpha\neq 0$. This indicates a non-singular smooth transition from contraction to expansion and complete turnaround of the universe. For $t>0$, we observe $\dot{a}(t)>0 $ which shows expansion after the bounce. In the vicinity of bounce, $\ddot{a }(t)>0 $ implies the universe expands at accelerating rate. This is in agreement with the concept of inflationary era. After a short period, it converts to $\ddot{a }(t)<0 $ shows decelerated expansion of the universe which provides explanation for matter dominated era.

\subsection{Hubble Parameter  ($HP$)}
We obtain its corresponding $HP$ value using standard relationship $H= \frac{\dot{a}}{a}$ as
\begin{equation}\label{15}
H = \frac{2\beta t}{(\alpha+\beta t^{2}) \log(\alpha+\beta t^{2})}.
\end{equation}
 The $HP$ in the contracting phase of the universe ($t<0$) is negative, $H< 0$. It becomes positive in post bounce expanding phase, i.e, $H>0$ for $t>0$ as shown in Fig. (\ref{h}). At the bounce $t=0$, the Hubble parameter vanishes, ($H=0$). The time derivative of $HP$ at bouncing point $\dot{H}=\frac{2\beta}{\alpha \log\alpha}$ is positive and finite. We note that for large values of parameter $\alpha$ i.e $\alpha\gg \beta$ or in the limit $\alpha\rightarrow\infty$ the term $\frac{2\beta}{\alpha \log\alpha}$ tends to zero, implying that $\dot{H}\rightarrow0$. As a result, $HP$ does not change its sign. This implies that the universe will follow its current state, either expanding or contracting at an approximately constant rate and no bouncing behaviour occurs under these conditions.
 \subsection{Deceleration Parameter ($DP$)}
\begin{figure}
\begin{minipage}{0.48\textwidth}
\centering
\includegraphics[width=8cm, height=7cm]{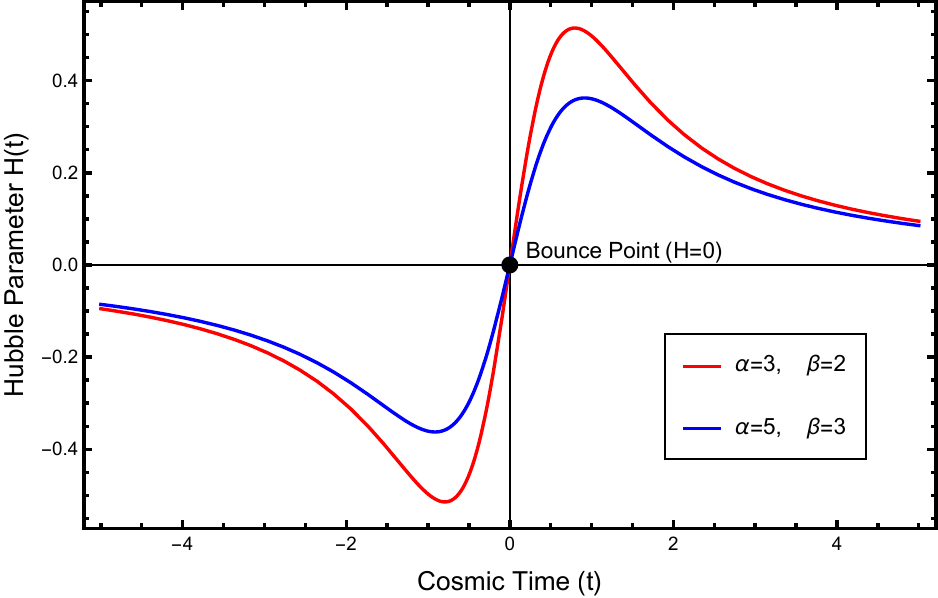}
\caption{ The plot of evolution of $HP$ with cosmic time by taking different values of model parameters $\alpha, \beta$.}\label{h}
\end{minipage}
\hfill
\begin{minipage}{0.48\textwidth}
\centering
\includegraphics[width=8cm, height=7cm]{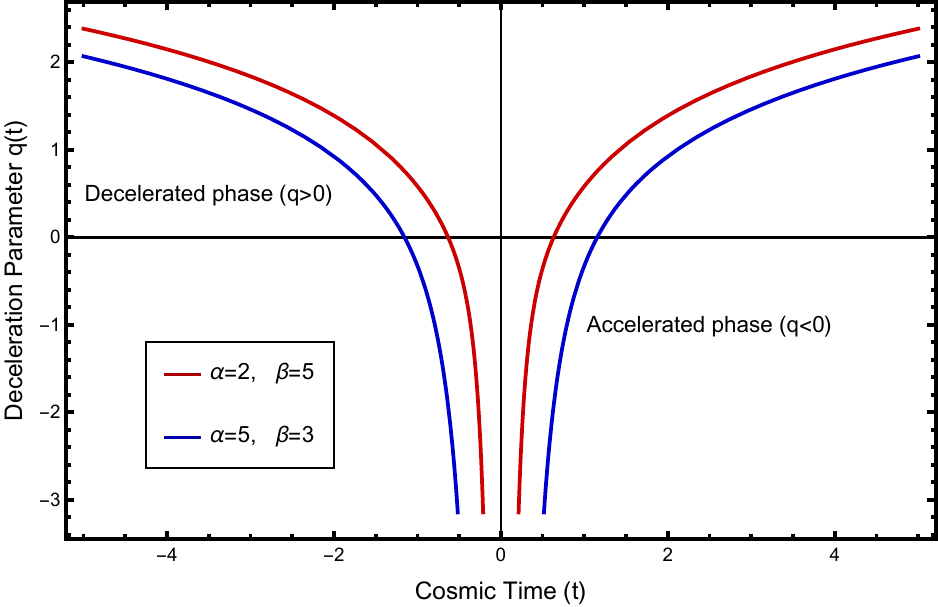}
\caption{The plot of evolution of $q$ with cosmic time by taking different values of model parameters $\alpha, \beta$ .}\label{q}
\end{minipage}
\end{figure}

Similarly, the $DP$ is defined as $q = - \frac{\ddot{a} a}{\dot{a}^2} =  -1 - \frac{\dot{H}}{H^2}$. We obtain the value of $DP$ as
\begin{equation}\label{16}
q= - \frac{1}{2}\left(\frac{\alpha}{\beta t^{2}}-1\right) \log(\alpha+\beta t^{2}).
\end{equation}
 The geometric parameter $q$ which is associated with the acceleration and deceleration of the universe behaves differently in transition phases. For positive value, $q>0$ corresponds to decelerated expansion, whereas a negative value $q<0 $ signifies an accelerated expansion of the universe. The left panel of the plot for $t<0$ in Fig. (\ref{q}) represents a decelerated contraction phase, followed by an accelerated contraction. We observe a strong accelerated contraction as $q<-1$ in the vicinity of the bouncing point. Since $q \propto \frac{1}{\dot{a}^2}$ and $\dot{a}(t) = 0 $ at $t=0$, $q$ diverges at bouncing point. After the bouncing epoch, $q$ rapidly evolves as $t>0$ and starts form negative values ($q<0$) implying an accelerated expansion just after the bounce. It is compatible with the inflationary era supposed to happen in very beginning of the universe. After a short time, DP takes positive values which represents the decelerated expansion compatible with matter dominated era.

\section{Dynamical Analysis of the Parameters and their Physical Interpretation}
\subsection{Energy Density ($ED$) and Pressure}
 The $ED$ and pressure of the universe are regulated by the matter, radiation and dark energy. From eq (\ref{13}), we calculated the expressions of $ED$ and pressure as
\begin{equation}\label{17}
\rho= \frac{2\beta[2\beta t^2(4\lambda + 3) + \lambda (-\alpha+ \beta t^2) \log(\alpha + \beta t^2)]}{(\lambda + 1)(2\lambda+1)[(\alpha + \beta t^2)  \log(\alpha + \beta t^2)]^2 }
\end{equation}
\begin{equation}\label{18}
p= -\frac{2\beta[2\beta t^2 + (3\lambda+2)(\alpha - \beta t^2)\log(\alpha + \beta t^2)]}{(\lambda + 1)(2\lambda+1)[(\alpha + \beta t^2)  \log(\alpha + \beta t^2)]^2}
\end{equation}
 It should be noted from eqs (\ref{17}) and (\ref{18}) that $\lambda$ must not take the values $-1,-\frac{1}{2}$ as these choices diverge both $ED$ and pressure ($\rho\rightarrow\infty$, $p\rightarrow\infty$) in all phases of cosmic evolution. During the contraction phase, the volume of the universe is shrinking, leading to an increase in $ED$ along with negative pressure due to compression of matter and radiation, since $ a(t)$ approaches its minimum value when $ t\rightarrow 0$. At $ t=t_{b}=0$, $\rho(t_{b})=-\frac{2\lambda\beta}{\alpha(\lambda+1)(2\lambda+1)\log\alpha}$ and as compared to singularity. The fact that the $ED$ is required to remain positive throughout the evolution in order to maintain the physical behaviour. It is evident from the expression of $\rho(t_{b})$ that positive $ED$ can be achieved by keeping the appropriate balance of the coupling parameter$\lambda$, allowing it to take some negative values since $\alpha>1$ and $\beta>0$. For a positive value of $\lambda$, $\rho(t_{b})$ is negative which is not physically valid. The conditions $(\lambda+1)>0\implies \lambda>-1$, $(2\lambda+1)>0 \implies \lambda>-\frac{1}{2}$ and $\lambda<0$ are required to observe $\rho(t_{b})>0$. Therefore, positive $\rho(t_{b})$ is strongly dependents on negative coupling parameter $\lambda$ and it should lie between $-\frac{1}{2}<\lambda<0$ according to above constraints. Another conditions to obtain $\rho(t_{b})>0$ are $(\lambda+1)<0\implies \lambda<-1$, $(2\lambda+1)<0 \implies \lambda<-\frac{1}{2}$ and $\lambda<0$. Now, the coupling parameter $\lambda$ should lie in $\lambda<-1$. Thus, to observe $\rho(t_{b})>0$, the coupling parameter $\lambda$ should lie in $-\frac{1}{2}<\lambda<0$ or $\lambda<-1$. Following the bounce, where $a(t)$  is increasing as $t>0$, the universe enters into expanding phase and  $\rho$ begins to decrease gradually with cosmic time and reaching at constant value. This behaviour reflects the dilution of energy due to expansion of space time curvature.
  
 The negative behaviour of pressure $p(t_b)=-\frac{2\beta(3\lambda+2)}{\alpha(\lambda+1)(2\lambda+1)\log\alpha}$ is necessary to oppose the gravitational collapse for avoiding singularity at $t=t_{b}$ and for driving the bounce. To maintain the negative pressure, an essential requirement for violating energy conditions and enabling transition, the $\frac{3\lambda+2}{(\lambda+1)(2\lambda+1)}$ must be positive since $\alpha>1$ and $\beta>0$. While a positive value of $\lambda$ easily satisfies the condition but it provides a negative value of ED, rendering it physically unacceptable. For a negative value of $\lambda$, such that any two terms in $\frac{3\lambda+2}{(\lambda+1)(2\lambda+1)}$ are negative and remaining one is positive or all three terms are positive, the condition is satisfied, i.e, $p(t_b)<0$. Thus, negative values of $\lambda$ are also consistent with the requirement of negative pressure. It is evident that $\frac{3\lambda+2}{(\lambda+1)(2\lambda+1)}$ is positive if $\lambda$ lies in $-\frac{1}{2}<\lambda<0$ or $-1<\lambda<-\frac{2}{3}$. It is important to note that when $\lambda$ falls in $-\frac{1}{2}<\lambda<0$, both the requirement $\rho(t_b)>0$ and $p(t_b)<0$ are satisfied. 
 
 The positivity of $ED$ and the requirement of negative pressure impose constraints on coupling parameter $\lambda$. In $f(R,T)$ gravity, the known feature of matter and geometry coupling modifies the effective $ED$ and pressure differently because both contain different combinations of $\lambda$ in numerator and denominator. Therefore, the fulfillment of conditions are realized with restrictions on $\lambda$, where both the conditions are simultaneously satisfied.

\begin{figure}
\begin{minipage}{0.48\textwidth}
\centering
\includegraphics[width=7.5cm, height=8cm]{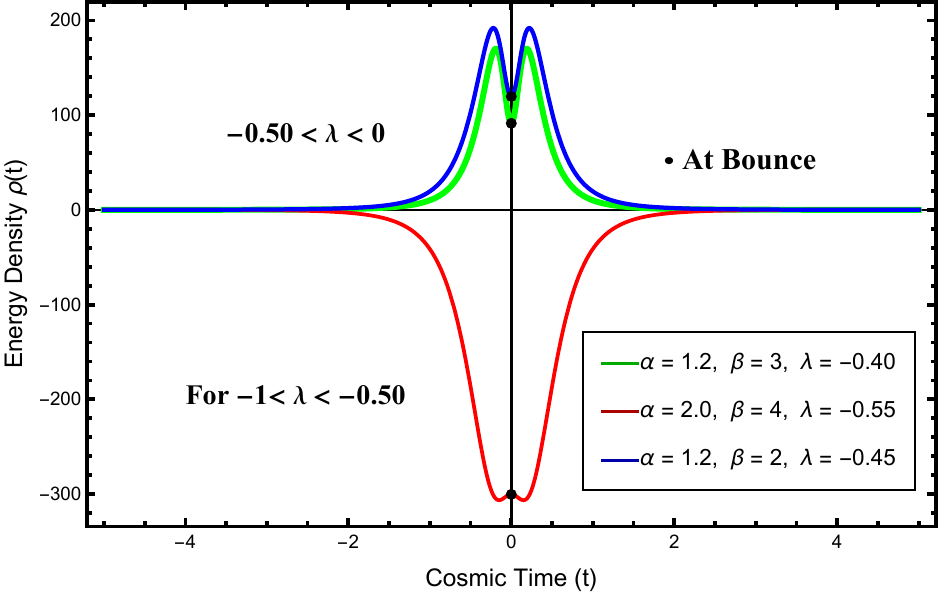}
\caption{The plot of evolution of $\rho$ with cosmic time for the constrained values of coupling parameter $\lambda$}
\end{minipage}
\begin{minipage}{0.48\textwidth}
\centering
\includegraphics[width=7.5cm, height=7.5cm]{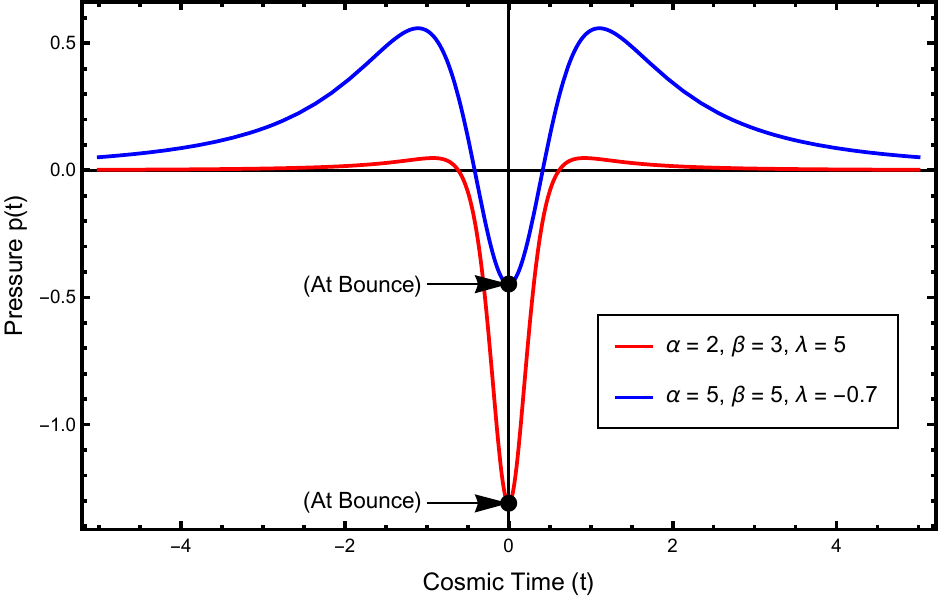}
\caption{The plot of evolution of pressure with cosmic time for the distinct values of model parameters $\alpha, \beta$ and positive value of coupling parameter$\lambda$}
\end{minipage}
\end{figure}
\subsection{Equation of State Parameter ($EoS$)}
Since $EoS$ parameter is defined as $w=\frac{p}{\rho}$  or  $w_{eff} =\frac{p_{eff}}{\rho_{eff}}$ , depends on the nature of $\rho ,  p $ and plays a crucial  role in governing the transition between different phases of the universe. Using eq. (\ref{17}) and (\ref{18}) we get
\begin{equation}\label{19}
w= \frac{-2\beta t^2 + (3\lambda+2)(-\alpha +\beta t^2)\log(\alpha + \beta t^2)}{2\beta t^2 (3+4\lambda) + \lambda(-\alpha + \beta t^2)  \log(\alpha + \beta t^2)}.
\end{equation}
We observe that $w>-1$ for both positive and negative time domains except in the vicinity of bounce. When $t\rightarrow0$, the model enters a phantom like phase that leads to the violation of NEC, which is necessary for non-singular transition from contraction to expansion. At bouncing point $t_b=0$, $w(t_{b})=\frac{3\lambda+2}{\lambda}$ and depends only on coupling parameter $\lambda$. We get $w(t_{b})=-1$ for $\lambda=-\frac{1}{2}$, $w(t_{b})=0$ for $\lambda=-\frac{2}{3}$ and $w(t_{b})=1$ for $\lambda=-1$. These values show that EoS parameter $w(t_{b})$ does not enter phantom region for a range of the model parameters. Notably, the curve of $w(t)$ splits itself for any value of coupling parameter at bouncing point. Afterward, the model gradually evolves from phantom regime and crosses the $\Lambda-CDM$ line ($w= -1$) and eventually approaches $w\approx 1$ which indicates a stiff matter dominated era associated with late time deceleration.
\begin{figure}
\centering
\includegraphics[width=7cm, height=7cm]{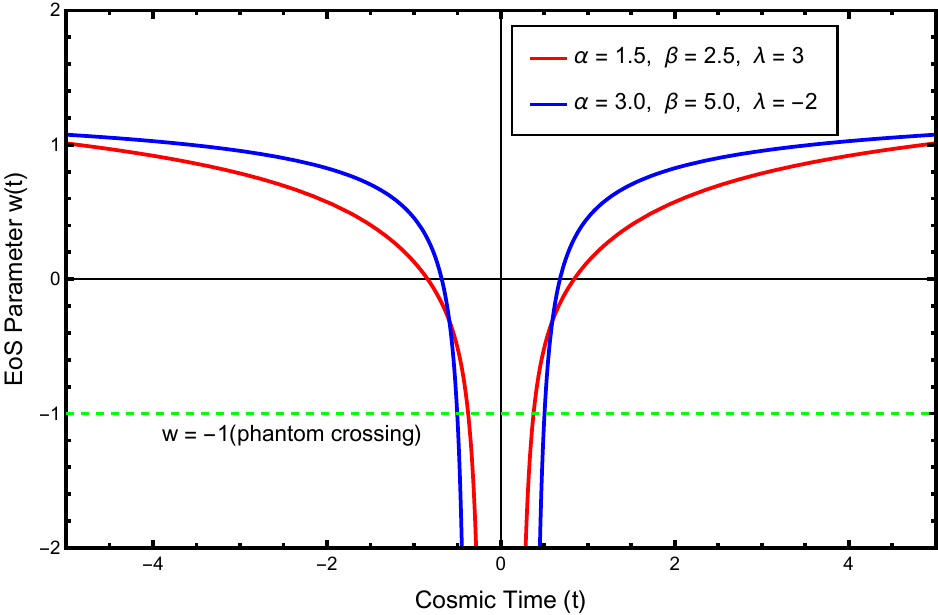}
\caption{ The plot of evolution of the $EoS$ parameter $w(t)$ with cosmic time for different values of parameters $\alpha$, $\beta$ and $\lambda$}\label{eos}
\end{figure}
\subsection{Energy conditions}
Energy conditions are the fundamental component, used to describe the physical behaviour of  geometric evolution of the spacetime. The effective density and pressure in energy conditions is defined as,
\begin{itemize}
\item Null Energy Condition (NEC): $\rightarrow \rho+p\geq0$,
\item Weak Energy Condition (WEC): $\rightarrow \rho\geq0$,
\item Strong Energy Condition (SEC): $\rightarrow \rho+3p\geq 0$
 \item Dominant Energy Condition (DEC): $\rightarrow \rho > \mid p\mid \geq0$
 \end{itemize}
In the framework of standard FLRW cosmology, the violation of the $NEC$ ($\rho+p<0$) corresponds to a positive time derivative of the HP, i.e., $\dot H>0$. However, in modified gravity theories, this correspondence must be interpreted in terms of effective ED and pressure. On combining eqs (\ref{17}) and (\ref{18}), we get
\begin{equation}\label{20}
\rho+p = \frac{ 4\beta[2\beta t^2 + (- \alpha+ \beta t^2) \log(\alpha+\beta t^2)]}{(\lambda+1)[(\alpha + \beta t^2)  \log(\alpha+\beta t^2)]^2}
\end{equation}
\begin{equation}\label{21}
\rho + 3p = \frac{4\beta[4\lambda \beta t^2 + (3+5\lambda)(-\alpha+\beta t^2)\log(\alpha+\beta t^2)]}{(\lambda+1)(2\lambda+1)[(\alpha + \beta t^2)  \log(\alpha+\beta t^2)]^2}
\end{equation}

\begin{figure}
\begin{minipage}{0.48\textwidth}
\centering
\includegraphics[width=7.5cm, height=7cm]{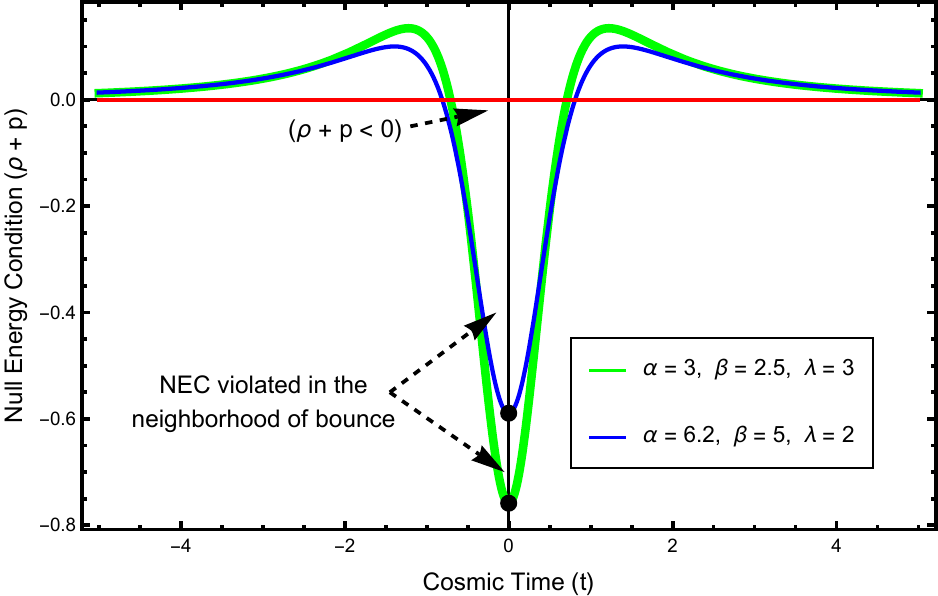}
\caption{The plot of NEC ($\rho+p$) with cosmic time for different values of parameters $\alpha, \beta$ and $\lambda$.}\label{nec}
\end{minipage}
\begin{minipage}{0.48\textwidth}
\centering
\includegraphics[width=7.5cm, height=7cm]{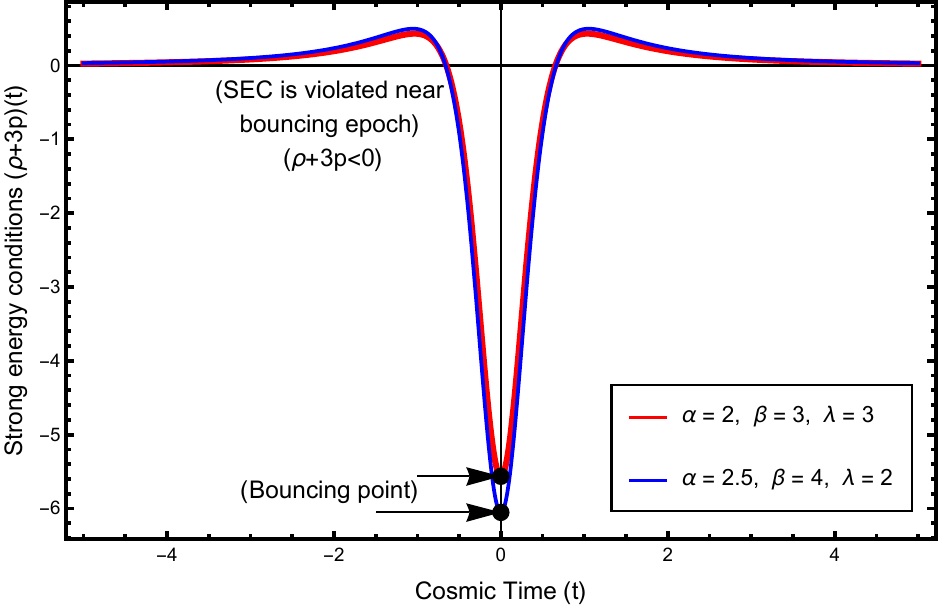}
\caption{The plot of SEC ($\rho+3p$) with cosmic time for different values of parameters $\alpha, \beta$ and $\lambda$.}\label{sec}
\end{minipage}
\end{figure}

 In the model $f(R,T)= R+\lambda T $, where $T=(\rho-3p)$, the matter content directly modifies spacetime geometry that naturally leads to the violations of the classical energy conditions \cite{consfunc,Houndjo a,Houndjo b}. This coupling makes the effective pressure negative enough but finite near the bouncing point to violate NEC, i.e, ( $\rho_{eff}+p_{eff}<0$). However, from eqs. (\ref{20}) and (\ref{21}), $\lambda\neq(-1,-\frac{1}{2})$ otherwise NEC and SEC will diverge. The graphical representation of violated $NEC$ and $SEC$ are shown in Figs. (\ref{nec}) and (\ref{sec}). Both $NEC$ ($\rho_{eff}+p_{eff}\geq0$) and $SEC$ ($\rho_{eff}+3p_{eff}\geq0$) are satisfied outside the bouncing regime which indicates the restoration of energy conditions. 
  
  At the bouncing point $t=0$, the null energy condition reduces to
  	\begin{equation}
  		\rho + p = -\frac{4\beta}{\alpha(\lambda+1)\log \alpha}.
  	\end{equation}
  	Since $\beta > 0$ and $\alpha > 1$, the sign of $(\rho + p)$ is determined entirely by the factor $(\lambda+1)$. The $NEC$ violation $(\rho + p < 0)$, therefore requires
  	$1 + \lambda > 0 \quad \Rightarrow \quad \lambda > -1$.
  	This shows that the coupling parameter $\lambda$ is not restricted to be strictly positive. Instead, both negative and positive values of $\lambda$ are allowed, provided $\lambda > -1$. 
  	The strong energy condition at the bounce point is given by
  	\begin{equation}
  		\rho + 3p = -\frac{4\beta(5\lambda+3)}{\alpha(\lambda+1)(2\lambda+1)\log \alpha}.
  	\end{equation}
  	  	The violation of the $SEC$ depends on the combined behavior of the numerator term $(5\lambda+3)$ and the denominator term $(\lambda+1)(2\lambda+1)$. This leads to a more restrictive condition on $\lambda$, which must be chosen such that the overall expression remains negative.
  	  	A detailed analysis shows that $SEC$ violation can be achieved when coupling parameter $\lambda$ falls in $-1 < \lambda < -\frac{3}{5}$ or $-\frac{1}{2} < \lambda$. It can be observed that violation of $NEC$ and $SEC$ at the bounce is achieved in a common region $-\frac{1}{2} < \lambda$. As discussed in $5.1$, when $\lambda$ falls in $-\frac{1}{2}<\lambda<0$ both the requirement $\rho(t_b)>0$ and $p(t_b)<0$. It is important to note that both $NEC$ and $SEC$ are also satisfied in the region. Thus, all the required conditions are satisfied in the region  $-\frac{1}{2}<\lambda<0$ at the bounce.
  	  	
These results demonstrate that the violation of energy conditions in the present model is not governed by the sign of $\lambda$ alone, but rather by its magnitude and its interplay with other model parameters. This provides a more refined and physically consistent understanding of the role of matter–geometry coupling in driving the bouncing dynamics.

\subsection{Stability Analysis of the Model}
\begin{figure}
\centering
\includegraphics[width=7cm, height=6cm]{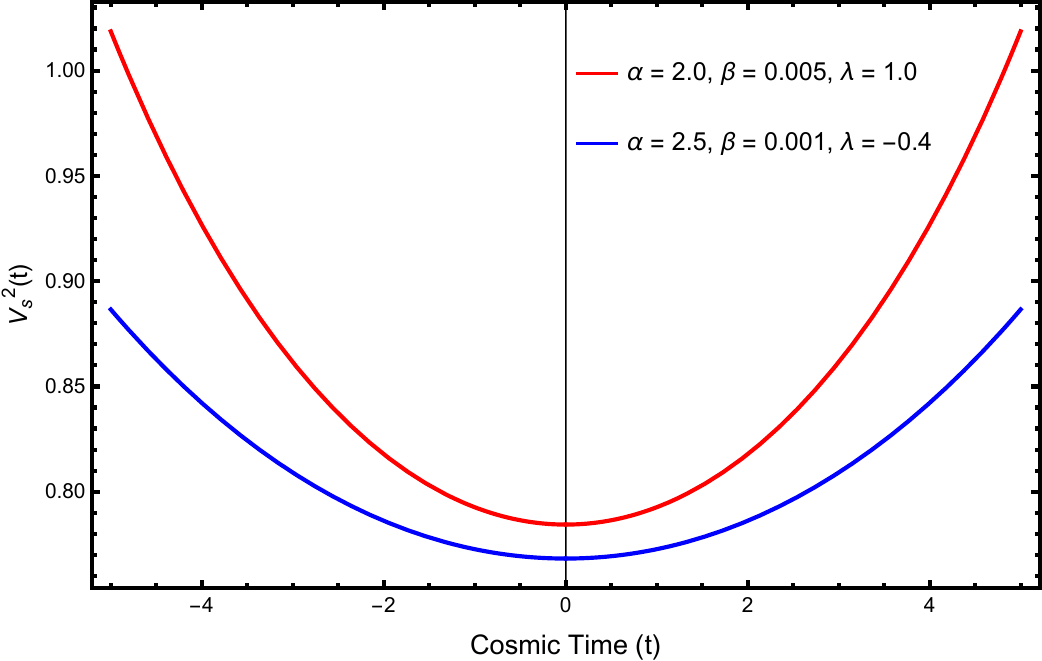}
\caption{The plot of stability analysis as speed of sound $V_{S}^2$ for different values of $\alpha$, $\beta$ and $\lambda$}\label{vs}
\end{figure}

The stability analysis is essential to validate the proposed bouncing model whether consistent with the behaviour of the real universe or not. In $f(R,T)$ theory such analysis becomes crucial because curvature scalar $R$ directly couples with  matter via trace $T$ of energy-momentum tensor. This coupling introduces additional force, and modifies conservation law ($ \nabla^{\mu} T_{\mu\nu} \neq 0 $) which makes the cosmological solutions sensitive to small peturbations. These perturbations can be generated by gravitational effects, compression of baryonic matter and quantum fluctuations in the comic fluid. In this work, we focus on the speed of sound as a diagnostic for stability \cite{Gs}. The sound speed determines how small perturbations in pressure and energy density propagate in the cosmic fluid, and defined as in the terms of derivative of effective energy density and pressure by $V_{s}^{2} = \frac{\dot{p}}{\dot{\rho}}$, which gives
\begin{widetext}
\begin{equation}\label{22}
V_{s}^{2}= \frac{- 4\beta t^2 - 3\lambda(\alpha - \beta t^2) \log  (\alpha+\beta t^2) + ( 3\lambda+2) (- 3\alpha + \beta t^2) \log^2(\alpha+\beta t^2)}{4\beta t^2 (4\lambda+3) - 3(3\lambda+2)(\alpha - \beta t^2)  \log (\alpha+\beta t^2) + \lambda (-3\alpha + \beta t^2) \log^2(\alpha+\beta t^2)}
\end{equation}
\end{widetext}
 The classically stability of the model requires that the squared sound speed remains positive, $V^{2}_{s}>0$. This ensures that small perturbations in the cosmic fluid do not grow exponentially. A careful inspection of Eq. (\ref{22}) reveals that the behavior of $V_s^2$ strongly depends on the model parameters $\alpha$, $\beta$, and $\lambda$. By appropriately considering these parameters, it is possible to ensure that the sound speed remains within the physically acceptable range $0 < V_s^2 \leq 1$ near the bouncing point and during the late-time evolution, see Fig \ref{vs}. 	
 	In particular, at the bouncing point $t=0$, the squared sound speed reduces to
 	\begin{equation}
 		V_s^2 (t_b) = \frac{(3\lambda+2)\log \alpha + \lambda}{\lambda \log \alpha + (3\lambda+2)},
 	\end{equation}
 	which remains positive and subluminal for suitable choices of $\lambda$ and $\alpha$. This confirms that the model can be made both stable and causally consistent in the physically relevant regime. Also, the curve remains positive for all positive values of $\lambda$ and does not depend on the value of $\alpha$. Specifically, for $\lambda = -1$ the propagation of sound remains unity ($V_{s}^{2}=1$), ensuring perturbations remain controlled and travel at most at the speed of light. On the other hand, when $V_{s}^{2}>1$, it do not affect classical stability, it violates causality conditions by allowing super luminal propagation of waves which is physically unacceptable within a causal relativistic framework.
\section{Thermodynamics}
In this section, we discuss the thermodynamic behaviour of our bouncing model. A deep connection between gravity and thermodynamics was first revealed in by suggesting that gravity may have a thermodynamic origin \cite{jac}. This idea is further strengthened by Padmanabhan \cite{pad} who showed that Einstein equation can be interpreted as thermodynamic identity. These works provide fundamental relation between spacetime geometry and thermodynamics laws and motivated further investigations in gravitational theories.

The generalized second law of thermodynamics ($GSL$), states that the total entropy of the universe, including the entropy of horizon and that of the matter enclosed by it, should be a non-decreasing function of cosmic time. The validity of $GSL$ has been widely examined in $GR$ as well as in various modified theories of gravity, where horizon thermodynamics treated as a powerful tool to test the viability of cosmological models. It is also interesting to note that thermodynamic considerations, where the model involves the transition from contracting phase to expanding phase, accompanied by violation of energy conditions.
The total entropy of the universe is defines as \cite{momeni}:
\begin{equation}\label{233}
	S_{total}=S_{h}+S_{in}+S_{prod}
\end{equation}
where $S_{total}$ represents the total entropy, $S_{h}$ is the horizon entropy, ${S}_{in}$ is the entropy of total fluid inside the horizon and $S_{prod}$ denotes the additional entropy production arising from the non-conservation in $f(R,T)$ gravity due to matter-geometry coupling. The time variation of total entropy is given by 
\begin{equation}\label{23}
	\dot{S}_{total}=\dot{S}_{h}+\dot{S}_{in}+\dot{S}_{prod}.
\end{equation}
 To satisfy the $GSL$, the model must hold the condition $\dot{S}_{total}\geq0$. 
 
\begin{figure} 
 \includegraphics[width=8.5cm, height=7cm]{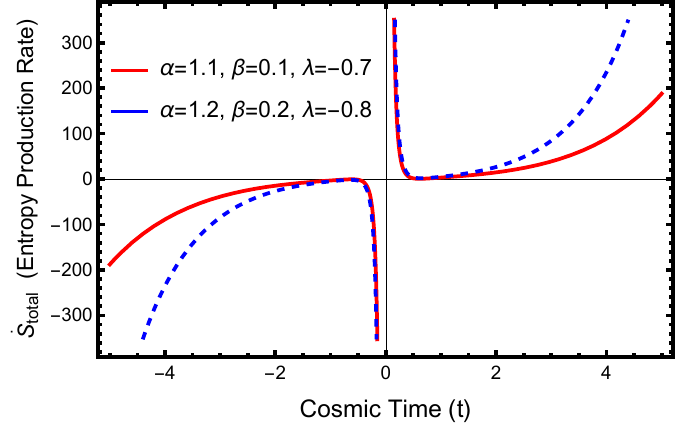}
 \caption{ The plot of $\dot{S}_{total}$ with cosmic time for different values of $\alpha$, $\beta$ and $\lambda$}.\label{en} 
 \end{figure}

For a flat FLRW universe, the radius of apparent horizon is given by $R_{h}=\frac{1}{H}$. The entropy associated with the apparent horizon is  $S_h = \frac{A}{4 G_{eff}}$, where $A$ is the horizon area defined as $A = 4 \pi R_h^2 = \frac{4 \pi}{H^2}$. In $f(R,T)$ gravity, the effective gravitational constant is $G_{eff}= \frac{G}{f_R}(1 + \frac{f_T}{8\pi G})$. We have considered $f(R,T) = R + \lambda T$ and $8\pi G=1$ in the model which imply $G_{eff}= \frac{1+\lambda}{8\pi}$. Therefore, $S_h = \frac{8\pi^2}{ (1+\lambda)H^2}$. The time variation of $S_h $ is obtained as
\begin{equation}\label{24}
 \dot S_h  = - \frac{16 \pi^2 \dot H}{(1+\lambda) H^3}
 \end{equation}
To obtain the entropy variation of the fluid inside the horizon, we apply the Gibbs equation
\begin{equation}\label{26}
T_{in} (dS_{in}+dS_{prod})=d(\rho_{(tot)}V_{h})+p_{(tot)}dV_{h}
\end{equation}
The term $V_{h}=\frac{4\pi}{3}R_{h}^3$ is the volume enclosed by the apparent horizon and $\rho_{(tot)}$ and $p_{(tot)}$ denote the effective total energy density and pressure of the cosmic fluid, including the contributions arising from gravitational coupling. Its rate change with time is calculated as
\begin{equation}\label{27}
\dot S_{in}+\dot{S}_{prod} =\frac{(\rho_{tot}+p_{tot})\dot V_h+V_h \dot \rho_{tot}}{T_{in}}.
\end{equation}
 The temperature of the fluid inside the horizon is denoted by $T_{in}$, which is assumed to be in thermal equilibrium with the apparent horizon, i.e., $T_{in} = T_h$. The temperature associated with the apparent horizon is taken to be the Hayward--Kodama temperature given by
\begin{equation}\label{28}
T_h = \frac{2H^2 + \dot H}{4\pi H},
\end{equation}
which reduces to the Hawking temperature $(T_H = \frac{H}{2\pi})$ in the de-Sitter limit, where $\dot H = 0$. Using the above relations, time variation in entropy of the fluid inside the apparent horizon can be expressed as
\begin{equation}\label{29}
\dot S_{in}+\dot{S}_{prod}=\frac{16\pi^2[6\dot H((\lambda+1)H^2+(2\lambda+1)\dot H)-\lambda H \ddot H]}{3(\lambda+1)(2\lambda+1)H^3(2H^2+\dot H)}.
\end{equation}
Hence, from eqs (\ref{23}), (\ref{24}) and (\ref{29}), we get
\begin{equation}\label{30}
	\dot{S}_{total}=-\frac{16\pi^2[6 \lambda \dot H H^2-3(2\lambda+1)\dot H^2+\lambda H \ddot H]}{3(\lambda+1)(2\lambda+1)H^3(2H^2+\dot H)}
\end{equation}
It is easy to observe that $\dot{S}_{total}$ diverges at the bouncing point as $H(t_b)=0$, also see Fig. \ref{en}.   Now, using the value of $H$ from eq (\ref{15}), we get the final value as
\begin{widetext}
\begin{equation}\label{31}
\begin{aligned}
\dot{S}_{total}=-\frac{
	\begin{aligned}
		&4\pi^2(\alpha + \beta t^2)\log(\alpha + \beta t^2)
		[4(10\lambda+3) \beta^2 t^4 +6 (5\lambda+2) \beta t^2 (-\alpha +  \beta t^2)\log(\alpha +  \beta t^2) \\
		& +(\log(\alpha + t^2 \beta))^2(-6 \alpha \beta (\lambda+1) t^2+(4\lambda+3) \beta^2 t^4+3 \alpha^2 (2\lambda+1))]
	\end{aligned}}{3 \beta^2 t^3 (\lambda+1) (2 \lambda+1) (-2  \beta t^2 + (-\alpha + \beta t^2)\log(\alpha + \beta t^2))}
\end{aligned}
\end{equation}
\end{widetext}

The behavior of the total entropy production rate is primarily dictated by the denominator, notably through the scaling $\dot{S}_{\text{total}} \propto t^{-3}$. In addition, the model parameter $\beta$ influences the evolution through time-dependent contributions such as $\beta t^2$ and $\beta t^3$, thereby controlling the relative dominance of different terms. For $t > 0$ (expanding phase), $\dot{S}_{\text{total}}$ becomes positive for suitable choices of model parameters, consistent with the GSL (see Fig.~\ref{en}). This reflects the standard thermodynamic arrow of time, where entropy increases as the universe expands away from the bounce.
For the same parameter choices, $\dot{S}_{\text{total}}$ becomes negative in the contracting phase ($t < 0$), which follows from the sign change induced by the $t^{-3}$ factor. This behavior is physically consistent and can be understood as the system evolving toward a minimum entropy configuration as the universe approaches the bounce. Notably, such a contracting phase is typically associated with the violation of the NEC, which is a generic requirement for the realization of a non-singular bouncing cosmology. Therefore, the decrease of entropy in the pre-bounce phase does not contradict fundamental thermodynamic principles but instead reflects the modified gravitational dynamics underlying the bounce. Although a reversed entropy evolution, i.e., $\dot{S}_{\text{total}} > 0$ for $t < 0$ and $\dot{S}_{\text{total}} < 0$ for $t > 0$, can arise for certain parameter choices, such behavior violates the thermodynamic arrow of time and is incompatible with the GSL in the post-bounce expanding phase. Hence, such scenarios are not physically admissible within the present framework.

It is important to note that the total entropy production rate $\dot{S}_{\text{total}}$ contains a factor proportional to $1/t^3$, which leads to a divergence at the bouncing point $t = 0$. This indicates that the GSL cannot be strictly defined at the exact moment of the bounce. This behavior can be physically understood as a consequence of the highly non-equilibrium nature of the bounce, where rapid changes in the Hubble parameter and the effective thermodynamic variables invalidate the assumptions of near-equilibrium thermodynamics on which the GSL is based.
 
 It is interesting to note the non-monotonic symmetric evolution of the entropy production rate in both contracting and expanding phases, the U-shape, as represented by dotted blue line in Fig. \ref{en}. In expanding phase, $t>0$, it starts with a very high value and decreases exponentially due to higher powers of time in denominator which suppress the entropy rate in the vicinity of the bounce. As a result, the entropy production rate initially decreases. As the universe evolves, the terms in the numerators start dominating which results in a almost constant evolution of entropy production rate and after some times it increases fast. Thus, we observe an almost U-shape evolution trajectories. The evolution in contracting phase is a mirror image of the entropy production rate in expanding phase. Therefore, the U-shaped behavior is a natural consequence of the competition between the time-dependent parameter denominator and the non-linear coupling terms of the model.    

\section{Consistency with Compact-Object Constraints}

The matter--geometry coupling parameter plays a fundamental role in the linear $f(R,T)$ gravity model and therefore its allowed range should be compatible with both cosmological and astrophysical observations. While the present work
constrains the coupling parameter from the requirement of obtaining a physically viable non-singular bouncing cosmology, compact astrophysical objects such as white dwarfs and neutron stars provide independent tests of the same matter--geometry coupling in the strong-field regime. Consequently, it is worthwhile to compare the cosmological constraint obtained in the
present work with the existing bounds derived from compact-object studies.

It should be emphasized that the compact-object analyses discussed below adopt the convention $f(R,T)=R+2\lambda_pT$,
whereas the present work considers $f(R,T)=R+\lambda T$.
These two formulations differ only by the normalization of the coupling parameter and are mathematically equivalent under the simple redefinition $\lambda=2\lambda_p$. Consequently, the constraint on the matter--geometry coupling parameter $\lambda$ required to realize a physically viable non-singular bouncing cosmology in the present work $-\frac{1}{2}<\lambda<0$,
corresponds to $-\frac{1}{4}<\lambda_p<0$
in the convention adopted by the compact-object literature. This allows a direct and meaningful comparison between the cosmological and astrophysical constraints.

In Ref.~\cite{gac}, the equilibrium configurations of white dwarfs were investigated within the framework of the linear $f(R,T)=R+2\lambda_pT$ gravity model. By comparing theoretical predictions with the available observational data, the authors obtained the lower bound $\lambda_p>-3\times10^{-4}$. Their analysis demonstrated that the matter--geometry coupling influences the mass, radius, pressure and energy density profiles of white dwarfs, thereby providing an independent
astrophysical constraint on the coupling parameter. Similarly, neutron stars were studied in Ref.~\cite{rl} using the same
functional form of $f(R,T)$ gravity together with realistic equations of state. By combining observations of massive pulsars with the gravitational-wave event GW170817, the authors concluded that physical consistency requires a small negative value of the coupling parameter and reported the bound $|\lambda_p|\lesssim0.02$. They further argued that the small sound speed in the neutron-star crust strongly restricts the allowed magnitude of the coupling parameter. Since
these constraints originate from a physical regime entirely different from cosmology, they provide an important complementary test of the theory.

Expressed in the same normalization convention, the cosmological constraint obtained in the present work, $-\frac{1}{4}<\lambda_p<0$, is broader than the compact-object interval reported in the literature. However, the two parameter spaces possess a non-empty overlap. Therefore,
the cosmological parameter space obtained in the present work is not inconsistent with the currently available compact-object constraints. Instead, the existing astrophysical studies may be regarded as selecting a physically preferred subset of the broader cosmologically allowed interval
required for the realization of a successful non-singular bounce.

The above comparison shows that the proposed bouncing cosmological model is compatible with the presently available compact-object constraints on the
matter--geometry coupling parameter after accounting for the different normalization conventions adopted in the literature. A comprehensive Bayesian analysis combining cosmological observations with compact-object
data would provide a more stringent determination of the coupling parameter. Such a joint analysis is beyond the scope of the present work and is left for future investigation.
\section{Discussion of Results}
In this section, we present a comprehensive discussion of the cosmological and thermodynamic implications of the logarithmic bouncing model constructed within the framework of $f(R,T)=R+\lambda T$ gravity. The analysis is organized by examining the behavior of key dynamical quantities—namely the scale factor, HP, DP, ED, pressure, EoS parameter, energy conditions, stability criteria, and thermodynamic variables across different cosmic phases. Particular emphasis is placed on the role of the matter--geometry coupling parameter $\lambda$, whose sign and magnitude critically regulate the realization of a non-singular bounce, the violation and restoration of energy conditions, classical stability, and the validity of the GSL. The results collectively demonstrate that a smooth transition from contraction to expansion can be achieved without encountering initial singularities, while maintaining physical viability within a constrained parameter space.

The logarithmic form of the scale factor $a(t)=m\log(\alpha+\beta t^2)$ with $m,\beta>0$ and $\alpha>1$ ensures a non-singular cosmological evolution. At the bouncing epoch $t=0$, the scale factor attains a finite minimum value $a_{\min}=m\log\alpha$, while $\dot{a}(0)=0$ and $\ddot{a}(0)>0$. These conditions guarantee a smooth transition from a contracting phase to an expanding phase, thereby replacing the big-bang singularity with a regular bounce. The parameters $\alpha$ and $\beta$ control the depth and sharpness of the bounce, respectively. 

The Hubble parameter $H(t)=\frac{2\beta t}{(\alpha+\beta t^2)\log(\alpha+\beta t^2)}$
exhibits the characteristic bounce signature. It remains negative during contraction, vanishes at the bounce ($H(t_b)=0$), and becomes positive in the expanding phase. Moreover, the time derivative at the bounce $	\dot{H}(0)=\frac{2\beta}{\alpha\log\alpha}>0$, confirms the violation of the NEC required for a successful bounce. Excessively large values of $\alpha$ suppress $\dot{H}$, indicating the sensitivity of the bounce to model parameters.

The deceleration parameter diverges at the bouncing point due to $\dot{a}=0$, which is a well-known feature of bouncing cosmologies. Prior to the bounce, the universe undergoes an accelerated contraction, while after the bounce it enters a decelerated expansion phase after a short accelerated expansion. This behavior is consistent with inflationary era followed by a matter dominated evolution. The ED and pressure derived from the modified field equations remain finite throughout the cosmic evolution, except for the excluded values $\lambda=-1$ and $\lambda=-\tfrac{1}{2}$. Positivity of the ED and negative pressure necessary to avoid gravitational collapse at the bounce commonly requires $-\frac{1}{2}<\lambda<0$.
 
The EoS parameter $w=p/\rho$ exhibits a phantom crossing near the bouncing epoch, allowing a temporary violation of the null energy condition. At the bounce $w_b=\frac{3\lambda+2}{\lambda}$, showing explicit dependence on the coupling parameter. After the bounce, $w$ smoothly evolves toward $\omega\simeq1$, indicating a stiff matter dominated decelerated expansion. Alongside, both the null and strong energy conditions are violated in the neighborhood of the bounce, which is essential for realizing a non-singular transition. Away from the bouncing epoch, these energy conditions are restored, ensuring consistency with standard cosmological behavior at late times. The classical stability of the model achieved at the bounce and throughout the contracting and expanding phases confirming the absence of growing perturbations. 

 Furthermore, the thermodynamic investigation reveals that entropy decreases during contracting phase and increasing during expanding phase. The decrease of entropy in the pre-bounce phase does not contradict fundamental thermodynamic principles but instead reflects the modified gravitational dynamics underlying the bounce. At the bouncing point, $\dot S_{tot}$ diverges. Physically, the transition form negative to positive values can be understood as a change in the direction of energy exchange between matter and geometry, consistent with the non-conservation of the energy--momentum tensor in $f(R,T)$ gravity. Near the bouncing epoch, this effective transition allows the universe to violate the null energy condition in a controlled manner while maintaining finite and stable cosmological quantities.

Another important outcome of the present work is that the cosmologically allowed range of the matter--geometry coupling parameter is found to be compatible with the currently available compact-object constraints. This result provides independent support for the physical viability of the proposed bouncing model by demonstrating that the parameter space required for the realization of a successful non-singular bounce is also consistent with astrophysical observations of compact objects. Therefore, the proposed bouncing model remains compatible with both cosmological and astrophysical constraints within the framework of linear $f(R,T)$ gravity.

\section{Conclusion}

In this work, a non-singular bouncing cosmological model has been developed within the framework of $f(R,T)=R+\lambda T$ gravity by adopting a logarithmic form of the scale factor. The proposed model successfully replaces the initial big-bang singularity with a smooth and finite bounce, characterized by a vanishing Hubble parameter and a positive time derivative at the bouncing epoch. The detailed dynamical analysis demonstrates that the universe undergoes a well-defined transition from a contracting phase to an expanding phase, accompanied by a temporary phantom regime that enables the necessary violation of energy conditions.

The matter--geometry coupling parameter $\lambda$ plays a crucial role in determining the physical viability of the model. The range $-\frac{1}{2}<\lambda<0$ is essential for positive energy density and negative pressure required to trigger the bounce and to ensure thermodynamic consistency during the expanding phase.

Furthermore, the stability analysis based on the squared sound speed confirms that the bouncing solution is classically stable, with perturbations remaining well-behaved across the bounce. The thermodynamic investigation reveals that the GSL remains valid throughout the post-bounce evolution. In the pre-bounce phase, entropy decrease toward a minimum entropy configuration as the universe approaches the bounce, which is a physically well behaved phenomenon. 

Finally, the cosmologically allowed range of the matter--geometry coupling
parameter is found to be compatible with the currently available compact-object constraints. This consistency further supports the physical
viability of the proposed non-singular bouncing model within the framework of linear $f(R,T)$ gravity.

Overall, the present study demonstrates that the logarithmic scale factor model in $f(R,T)$ gravity provides a consistent, stable, and thermodynamically viable framework for describing non-singular bouncing cosmologies while remaining compatible with the currently available compact-object constraints.
\section*{Declaration of competing interest} The authors declare that they have no known competing financial interests or personal relationships that could have influenced the work reported in this study.

\section*{Acknowledgments}
The authors are thankful to the anonymous reviewers for constructive comments, which helped to improve the paper's quality in its present form. The authors S.H. Shekh and Pankaj Kumar appreciate the help and resources given by the IUCAA Pune, India.

\end{document}